\documentclass{article}

\usepackage{microtype}
\usepackage{graphicx}
\usepackage{booktabs} 
\usepackage[frozencache=true]{minted}
\usepackage{xcolor}
\usepackage{flushend}
\usepackage{listings}
\usepackage[OT1]{fontenc}
\usepackage{subcaption}

\usepackage{enumitem}      
\usepackage[compact]{titlesec}
\usepackage{rotating} 

\definecolor{light-gray}{rgb}{0.99,0.99,0.99}
\setminted{bgcolor=light-gray}  
\setminted{fontsize=\footnotesize}

\setlist{noitemsep,topsep=0pt,partopsep=0pt}

\PassOptionsToPackage{hyphens}{url}\usepackage{hyperref}

\usepackage[anythingbreaks]{breakurl}

\newcommand{\code}{\mintinline[fontsize=\footnotesize]{python}}
\newcommand{\mintedsize}{\footnotesize}

\usepackage[accepted]{sysml2019}

\sysmltitlerunning{AutoGraph: Imperative-style Coding with Graph-based Performance}

\begin{document}

\twocolumn[
\sysmltitle{AutoGraph: Imperative-style Coding with Graph-based Performance}



\sysmlsetsymbol{equal}{*}

\begin{sysmlauthorlist}
\sysmlauthor{Dan Moldovan}{goog}
\sysmlauthor{James M Decker}{purd}
\sysmlauthor{Fei Wang}{purd}
\sysmlauthor{Andrew A Johnson}{goog}
\sysmlauthor{Brian K Lee}{goog}
\sysmlauthor{Zachary Nado}{goog}
\sysmlauthor{D Sculley}{goog}
\sysmlauthor{Tiark Rompf}{purd}
\sysmlauthor{Alexander B Wiltschko}{goog}
\end{sysmlauthorlist}

\sysmlaffiliation{goog}{Google Brain, Cambridge, MA}
\sysmlaffiliation{purd}{Purdue University}
\sysmlcorrespondingauthor{Tiark Rompf}{tiark@purdue.edu}
\sysmlcorrespondingauthor{Alexander B Wiltschko}{alexbw@google.com}

\sysmlkeywords{Machine Learning, SysML}

\begin{abstract}

There is a perceived trade-off between machine learning code that is easy to write, and machine learning code that is scalable or fast to execute. In machine learning, {\em imperative} style libraries like Autograd and PyTorch are easy to write, but suffer from high interpretive overhead and are not easily deployable in production or mobile settings. {\em Graph-based} libraries like TensorFlow and Theano benefit from whole-program optimization and can be deployed broadly, but make expressing complex models more cumbersome. We describe how the use of staged programming in Python, via source code transformation, offers a midpoint between these two library design patterns, capturing the benefits of both. A key insight is to delay all type-dependent decisions until runtime, similar to dynamic dispatch. We instantiate these principles in AutoGraph, a software system that improves the programming experience of the TensorFlow library, and demonstrate usability improvements with no loss in performance compared to native TensorFlow graphs. We also show that our system is backend agnostic, targeting an alternate IR with characteristics not found in TensorFlow graphs.


\end{abstract}
]

\printAffiliationsAndNotice{}

\section{Programming Paradigms for Machine Learning}


Programming platforms specialized for machine learning (ML) are undergoing widespread adoption, as ML models such as neural networks demonstrate state-of-the-art performance on many important industrial problems like translation and image recognition. 
In order to support this proliferation of use, there has been rapid development of
platforms for building new ML models. These platforms follow
two main paradigms, {\em graph-based} programming and
{\em imperative} programming. These have also been labeled {\em Define-and-run} and {\em Define-by-run} \cite{tokui2015chainer}.


{\em Graph-based} systems like TensorFlow and Theano use a high-level language (typically Python) to metaprogram a lower-level intermediate representation (IR) of computation \cite{abadi2016tensorflow, al2016theano}. In TensorFlow's case, this IR provides a representation that can then be automatically distributed across a datacenter, executed on accelerator hardware like GPUs or TPUs, deployed to mobile devices or web servers, and can benefit from whole-program optimization. The computational gains are significant, but come at the cost of additional cognitive load for developers.

{\em Imperative} programming systems like PyTorch and Autograd \cite{paszke2017automatic, maclaurin2015autograd} run user code directly, building up a representation of the user's program incrementally for either automatic differentiation or compilation. TensorFlow also supports imperative-style coding via ``eager execution'', where user-written Python code immediately executes TensorFlow kernels, without a graph being built. Such systems allow the user to enjoy the benefit of traditional imperative coding, but have reduced opportunities for program optimization, scalable computation and portability.

The differences between these approaches are especially apparent for models that require data-dependent control flow, such as conditionals or loops, which are important for state of the art methods in Reinforcement Learning, sequence-based models, and many other emerging research areas. {\em Imperative} platforms allow a user to write idiomatic and native Python control flow, using traditional syntax for data-dependent control flow operations such as conditionals and loops. However, this approach reduces opportunities for whole-program optimization and requires retracing on every execution for automatic differentiation. {\em Graph-based} platforms avoid this issue, but do not allow traditional Python syntax for data-dependent control flown, and instead require any data-dependent control flow to be expressed in a functional form. This is required because Python does not natively support deferring the execution of control flow.

While {\em graph-based} and {\em imperative} programming are often presented as orthogonal and independent programming paradigms, we provide an approach that offers the best of both, retaining {\em imperative} usability benefits while still yielding {\em graph-based} performance and portability benefits.
We note that this approach assumes the ability to transform code into a specialized IR, and that this IR confers real benefits to the programmer such as speed, memory and numerical stability optimizations, as well as deployability to a variety of platforms. However, like many IRs, we also assume that it is cumbersome to program directly. Due to its widespread usage and robust IR, we focus much of our discussion on TensorFlow graphs, but show in our evaluation (Section~\ref{treelstm}) that this approach is completely independent of any back-end, and indeed, we can represent programs not easily expressible in TensorFlow's IR by selecting a different back-end for our code generation engine to target.

The contributions of this paper are as follows:

\begin{itemize}
    \item We propose a new methodology that provides users the expressive power of {\em imperative} ML systems, while retaining the performance and portability of {\em graph-based} systems. 
    \item We demonstrate this methodology in Python using static analyses and {\em source code transformations} (SCT).
    \item Using these analyses and code transforms, we enable staged programming in Python dispatching on runtime type information, in most cases requiring no additional annotations.
    \item We use our system, called AutoGraph, to convert idiomatic Python into TensorFlow Graph IR. We show that AutoGraph generalizes to target other back-ends, and can convert Python code into the Lantern IR, which supports features absent from the TensorFlow Graph IR, such as re-entrant function calls.
    \item We demonstrate that our system allows a user to easily express complex ML programs that lower to an optimized IR, and run as fast as hand-written alternatives.
\end{itemize}

\section{Related Work}
A number of existing systems and approaches also aim to provide an easy-to-use programming interface for defining ML models without degrading performance. One such example is the Open Neural-Network eXchange (ONNX) format \cite{onnx2018}, which provides an IR with APIs for many high-level front-ends that can target a number of popular back-ends focused on optimization and high-performance computing. This IR is exhibited as a computation graph, generated through the use of tracing, as in many {\em imperative} systems. ONNX provides insight into the ability to use an IR as the broker between {\em imperative} and {\em graph-based} systems, though extracting graphs via tracing may yield a loss of control flow information due to the inability to capture data-dependent control flow.

Another recent approach is that of PyTorch's Torch Script framework \cite{torchscript2018}. While based on Python AST translation similar to AutoGraph, there are a number of important differences, most notably the lack of staging beyond shape propagation on a dynamically-shaped graph. A more complete comparison of Torch Script and AutoGraph can be found in Section~\ref{torchscript_ag}. The Myia system \cite{merrienboer2018myia} provides a similar facility Torch Script, where the user expresses numeric code in Python which is then parsed into a graph-based IR distinct from the Python AST. J\textsc{ANUS}~\cite{Janus} operates like a JIT compiler from Python bytecode to TensorFlow graph code, modifying the Python interpreter. In contrast, AutoGraph works as a stand-alone library performing source-to-source transformations.

Providing easier deferred execution using staged programming or multiple dispatch has a long history. Notable examples include Lightweight Modular Staging's type-based deferred execution model \cite{rompf2010lightweight}, the paired use of Lua and Terra to stage high-performance numerical code \cite{devito2013terra}, and Julia's multiple dispatch system \cite{bezanson2012julia}. Libraries implementing or using code rewriting in Python have been in limited use, including the privacy- and confidentiality-aware Jeeves system \cite{YangHASFC16}, which relies on MacroPy \cite{macropy}, as well as the Hy system, a Lisp dialect embedded in Python \cite{hy}. However, each of these approaches alone, without substantial modification, are inappropriate for the Python language.

Other efforts contributed a variety of ML frameworks with different features. Lantern \cite{wang2018a, DBLP:journals/corr/abs-1803-10228} applied concepts of programming languages research (delimited continuations and multi-stage programming) to implement an expressive {\em graph-based} ML framework. Tangent \cite{Tangent} performs automatic differentiation using SCT. Dynet \cite{dynet} is a define-by-run system with a dynamic batching runtime for automated batching computations. MXNet \cite{DBLP:journals/corr/ChenLLLWWXXZZ15} offers both options of {\em define-by-run} and {\em graph-based} through the use of different syntax. Both chainer \cite{chainer_learningsys2015} and torch-autograd, a Lua port of the Autograd library \cite{torch-autograd} are pure define-by-run systems. Numba \cite{Lam:2015:NLP:2833157.2833162} translates annotated Python functions to machine code at runtime.

\section{Programming TensorFlow}
\label{ProgrammingTF}
The TensorFlow software programming system has become popular for ML practitioners, particularly those focusing on large-scale training and deployment \cite{tensorflow-is-used-a-lot}.
ML programs naturally execute in separate stages, as model architecture and data examples become available at different points in a program's lifecycle, and TensorFlow makes these stages explicit. A TensorFlow user must first build up a representation of the computation to be run, and then later in their program, specify that the computation should be executed.
Dataflow graphs are used for this representation, because they can be readily optimized, distributed and deployed.
This programming model is sometimes non-obvious, leading to difficult usability issues and bugs, and is particularly acute in the case of specifying control flow.
For example, some control flow constructs should be included in the lowered IR, while others are meant to specify whether or not computation should be staged into the IR. A common coding pattern is to conditionally stage computation using model hyperparameters:

\begin{listing}[ht]
\begin{minted}[fontsize=\mintedsize]{python}
# Conditional on bool not added to graph
if HParams.nonlin == 'relu':
    x = tf.nn.relu(x)
else:
    x = tf.nn.tanh(x)
\end{minted}
\end{listing}

However, other uses of control flow are meant to be executed in a data-dependent manner:

\begin{listing}[ht]
\begin{minted}[fontsize=\mintedsize]{python}
# Conditional on Tensor added to graph
x = tf.cond(tf.reduce_sum(x) > 0,
  lambda: x * x, lambda: x)
\end{minted}
\end{listing}

In the code above, the conditional statement is expressed in a functional style so that it can be executed in-graph in a data-dependent manner. However, this clashes aesthetically and pragmatically with the imperative style of Python. This difficulty is exacerbated when the user needs to nest control flow, or use other Python idioms like \code{continue} and \code{break}. We would instead prefer to write

\begin{listing}[ht]
\begin{minted}[fontsize=\mintedsize]{python}
# Conditional on Tensor - staged
if tf.reduce_sum(x) > 0:
  x = x * x
\end{minted}
\end{listing}
  
and have it be automatically converted into the functional style. We  want this conversion to only occur for expressions using numeric types. Conditionals switching on plain Python booleans (e.g., the hyperparameter example above) should be executed imperatively, without staging.

\section{Extending Operator Overloading}
In the case of TensorFlow, metaprogramming dataflow graphs can be difficult for complex programs, but it is made easier via {\em operator overloading}. For example, the user does not need to type out \code{tf.add(a, b)}, but instead can simply use \code{a + b}. This is possible due to Python's ability to allow the programmer to overload a subset of the language. 
Python's approach to operator overloading allows custom classes, like the \code{Tensor} type in TensorFlow, to override some default functionality, like their behavior when used in binary operators (e.g. \code{+,*,-,\%,/,^,~}) or item access.\footnote{See Python Language Reference (\url{https://docs.python.org/3/reference/}), Section 3.3.}  

\begin{minted}[fontsize=\mintedsize]{python}
# Because Python lets us write this ...
class Tensor(_TensorLike):
  def __add__(self, right):
    return tf.add(self, right)

# ... we can write this
import tensorflow as tf
a = tf.constant(3)
b = tf.constant(4)
c = a + b
\end{minted}

This is a powerful facility in the Python language, but it unfortunately only extends to methods of objects or classes, and does not include programming constructs required to build modern ML models. For example, the behavior of conditionals cannot be overloaded in Python.

\begin{listing}[ht]
\begin{minted}[fontsize=\mintedsize]{python}
# We can write if statements...
if cond:
  ans = true_fn()
else:
  ans = false_fn()

# ... but we cannot overload them
def __if__(self, cond, true_fn, false_fn):
  if cond:
    return true_fn()
  else:
    return false_fn()
\end{minted}
\end{listing}

If overloading control flow syntax were possible, {\em imperative} programs would be able to generate full representations of user code, including previously-invisible loop and conditional statements. {\em Graph-based} programs would not need to require users to write their program control flow in a cumbersome functional form, because they could provide non-standard overrides of \code{__if__}, \code{__for__} and \code{__while__} and other useful parts of the Python language. 

To circumvent this limitation, we use SCT on whole functions to enable overloading non-local parts of the Python language.
We describe a specific instantiation of this system, called {\em AutoGraph} which uses SCT to allow users to target a lower-level IR while still writing idiomatic Python.

\section{Staged Programming for Real-World ML Systems}
Using the ability to overload arbitrary Python syntax, we built a staged programming system called AutoGraph for improving the performance of {\em imperative}-style ML programs and conversely, the simplicity of {\em graph-based} ML programs.

AutoGraph allows users to program using idiomatic and imperative-style Python, but still benefit from the advantages of TensorFlow graphs, and is exposed to users via a single-function API, as a Python function decorator as seen in Listing~\ref{lst:conversion}.

\begin{listing}[ht]
\begin{minted}[fontsize=\mintedsize]{python}
import autograph as ag

# AutoGraph converts whole
# functions via a decorator...
@ag.convert()
def f(x):
if x > 0:
  x = x * x
return x

# ... into a form where control flow
# and other idioms are overloadable
def new_f(x):
  def if_true():
    x_1 = x
    x_1 = x_1 * x_1
    return x_1
  def if_false():
    return x
  x = ag.if_stmt(
      ag.gt_(x, 0), if_true, if_false)
    return x
\end{minted}
\caption{AutoGraph automatically converts the code on the top into the code on the bottom (simplified example).}
\label{lst:conversion}
\end{listing}

AutoGraph works with control flow, such as \code{if}, \code{for} and \code{while} statements, even if they are arbitrarily nested or contain \code{break} and \code{continue} statements. 

The AutoGraph system can overload conditionals and loops via SCT, allowing us to deviate from Python's default behavior. Note that, using the same style of SCT, we may choose to overload some statements while preserving Python semantics for others. Because of this, we anticipate that this might be a tool of general interest to Python developers, or a feature that new language implementations might want to consider including. In order to transparently support control flow that is meant to either be staged or unstaged in TensorFlow, as in the conditional examples in Section~\ref{ProgrammingTF}, we must change the behavior of \code{if} statements based on the type of the boolean predicate.

\section{``Dynamic Dispatch" Enables Staged Programming in Python}
\label{dispatch}
Given that we can enable overloadable control flow in Python, we can redefine its default behavior by writing a non-default implementation of \code{ag.if_stmt}. In the case that a Python boolean is used as the predicate of a conditional, we would want to execute the conditional with normal semantics. However, if a TensorFlow Tensor is supplied, or some other specialized numeric type, we would want to stage more specialized code. A simplified version of \code{ag.if_stmt} is shown in Listing~\ref{lst:if}.

\begin{listing}[h]
\begin{minted}[fontsize=\mintedsize]{python}
def if_stmt(cond, body, orelse):
  if is_tensor(cond):
    return tf.cond(cond, body, orelse)
  elif cond:
    return body()
  else:
    return orelse()
\end{minted}
\caption{Simplified version of AutoGraph's conditional statement override.}
\label{lst:if}
\end{listing}

We use the term \emph{dynamic dispatch} to describe this runtime decision making, as it is analogous to dynamic \emph{method} dispatch common in object oriented programming. Dynamic dispatch critically allows us to seamlessly switch between two common uses of control flow in ML code -- a ``macro-programming" mode that switches or loops on the value of hyperparameters, and a data-dependent mode, where the control flow is lowered into the target IR. 

The same logic is applied to \code{for} and \code{while} loops in the equivalent of \code{ag.for_stmt} and \code{ag.while_stmt} functions. We also provide functionality for overriding the \code{print} statement, which is ordinarily incompatible with TensorFlow graphs, since \code{print} would log information immediately, and we instead want to log values at graph runtime.

Note that some native Python constructs, like \code{break} and \code{continue} statements have no direct representation in TensorFlow. This requires code transformations which entirely remove these statements without affecting program semantics. This is achieved by lowering the respective statements into equivalent TensorFlow constructs. For example, \code{continue} is lowered using extra variables and conditionals.
  
The dynamic dispatch approach incurs extra runtime overhead. Indeed, if AutoGraph was used to perform normal unstaged Python computation, it would be slower. However, because we target a lower-level IR that can be executed separately from the Python runtime, this overhead is amortized.


\paragraph{General Approach}


The conversion of a function proceeds as follows:

\begin{enumerate}
  \item Read the source code of the function and obtain its closure variables, if they are available. 
  \item Parse the source code into a Python AST, abstracting away small differences between Python versions.
  \item Transform the source code in multiple passes, with each pass consisting of two major steps:
  \begin{enumerate}
    \item Static analysis, detailed below. The AST is annotated with additional information that the actual transformation may use.
    \item AST transformations, where each transform handles a specific Python idiom. The specific transformations are detailed below.
  \end{enumerate}
  \item Serialize the final AST into output code.
  \item Load the new output code in as a Python function, and dynamically attach symbols corresponding to the original function's closure variables.
\end{enumerate}

\paragraph{Comparison with Static Methods}

It is possible to extract computation graphs from Python code statically, but doing so requires a strict set of constraints. Systems like Torch Script~\cite{torchscript2018} elect to impose these constraints in the form of a DSL which is a limited subset of Python. A major design decision in AutoGraph, however, is to allow users access to as much of the original Python interface as is possible (we discuss limitations to this in Section~\ref{discussion}). Furthermore, due to binding-time analysis, relying wholly on static methods disallows staged programming in Python without requiring some form of an ersatz static type system (e.g., static type annotations). While enabling staged programming in a dynamic setting for arbitrary types does require careful consideration~\cite{snek}, our decision to primarily target TensorFlow as a back-end significantly alleviates some of the implementation pains due to the central focus of an array-based type (tensor). We discuss this in detail in Section~\ref{conversion}.

\section{Code Analyses and Conversion}
\label{conversion}
Only a subset of Python can be trivially converted, and substantial rewriting of user-provided code is necessary to enable the overloading required for staged programming. For example, loops and conditionals need to be rewritten in functional form; nonlocal control flow statements need to be lowered. We perform these rewrites with the aid of dataflow and other analyses of program structure. We also separate these rewrites into multiple specialized passes.
 
\subsection{Dataflow Analysis}
Each specialized pass is preceded by several dataflow analysis passes. These are described below, in the order that they are run.

\paragraph{Control Flow Graph Construction}
A standard intra-procedural control flow graph (CFG) supports several static analyses.



\paragraph{Qualified Name Resolution}
We create the abstraction of qualified names to extend the notion of symbols to include compound names such as \code{a.b}. For example, the qualified name \code{a.b} roughly corresponds to the AST: \code{Attribute(name=Name('a'), attr='b')}. 

\paragraph{Activity Analysis}
Here we annotate AST nodes with the list of symbols read and modified by the respective statement. Only direct modifications are considered writes. For example, in the statement \code{a.b = c}, \code{a.b} is considered to be modified, but \code{a} is not. 
The activity analysis also keeps track of lexical scopes, their nesting relationships (e.g. the parent scope) and the symbols they include.

\paragraph{Reaching Definitions Analysis}
This standard dataflow analysis annotates help identify the definition that reaches each name. Additionally, the list of symbols defined on entry of certain statements is also annotated.

\paragraph{Liveness Analysis}
This standard dataflow analysis identifies symbols that are live upon entry into or exit from certain statements, including compound statements like conditionals.

\subsection{Code Conversion Passes}
AutoGraph performs code conversion using an extensible system of multiple, typically independent, AST conversion passes. For example, one conversion pass rewrites the \code{if} statements into an overloadable functional form. Another pass lowers the \code{break} statements into new loop predicates and extra conditionals. This mechanism facilitates adding support for more Python idioms in time.

Currently, the transformations include the following, in order of application:

\paragraph{Directives}
Identifies calls to specific functions that serve as AutoGraph compilation directives and annotates the relevant AST nodes. An example of such a directive is \code{ag.set_loop_options}.
  
\paragraph{Break, Continue and Return Statements}
These are actually three separate passes, but are very similar in nature. In each case, the corresponding statement is lowered into conditionals or expanded loop conditions.

\begin{listing}[ht]
\begin{minted}[fontsize=\mintedsize]{python}
# Before conversion
if cond:
  return f(x)
return g(x)

# After conversion
if cond:
  return_value = f(x)
else:
  return_value = g(x)
return return_value
\end{minted}
\end{listing}





\paragraph{Assert Statements}
These are converted in-place to overloadable functional form.

\paragraph{Lists}
List idioms, including list literals and the \code{append} and \code{pop} function calls are overloaded with custom functions (e.g. \code{ag.list_append} and \code{ag.list_pop}) that allow staging the respective operation.



Array computations require an additional idiom not present in the standard Python library: the stack operation. AutoGraph provides the \code{ag.stack} function which can be overloaded in a manner consistent with the other overloads. Note that list access (e.g. \code{l[i]}) and mutation are deferred to a separate conversion pass which covers slice operators.

\paragraph{Slices}
Python does allow overloading the slice operators (\code{__setitem__}, \code{__getitem__}) in user classes. However, the slice write operation has the semantic of mutating the target. We rewrite slice writes to use value semantics as currently required by TensorFlow. For instance, \code{x[i] = y} is converted in-place to \code{x = ag.setitem(x, i, y)}. Slice read operations are converted mechanically.

\paragraph{Function Calls}
All function calls are overloaded. The overload will either dynamically convert the target function, call it as-is or replace it with a new function, depending on the characteristics of the function being called and the configuration of the conversion. For example, the built-in function \code{print} may be converted to \code{tf.print} (see Appendix~\ref{appendix:features} for details).

\begin{listing}[ht]
\begin{minted}[fontsize=\mintedsize]{python}
# Before conversion
def f(a, x):
  return a(x)

# After conversion (simplified)
def f(a, x):
  return ag.converted_call(a, x)
\end{minted}
\end{listing}


\paragraph{Control Flow}
This conversion pass replaces all local control flow with an overloadable equivalent functional form.

The \code{if} statement is stateless, therefore its functional form can be expressed using niladic functions that return all the variables modified inside the statement.

\begin{listing}[ht]
\begin{minted}[fontsize=\mintedsize]{python}
# Before conversion
if x > 0:
  x = x * x
  
# After conversion (simplified)
def true_fn():
  return x * x
def false_fn():
  return x
x = ag.if_stmt(x > 0, true_fn, false_fn)
\end{minted}
\end{listing}

Note that Python allows to define (i.e., assign for the first time) symbols inside the body of control flow statements and use them later. It is possible to write code where symbols may be undefined based on whether the branch of a conditional executed or not. However, the functional version of the conditional operators always sets the symbols that the conditional may modify in either branch. To simulate the undefined semantics, we use a special value to reify the ``undefined" state of a variable. This currently deviates from Python semantics, but we plan to remedy this by verifying and explicitly deleting ``undefined" symbols before they are used. 

The \code{while} and \code{for} loops are stateful, and their functional form requires functions whose arguments and return values represent the variables modified inside the loop (its state).

\begin{listing}[ht]
\begin{minted}[fontsize=\mintedsize]{python}
# Before conversion
while x > eps:
  x = f(x)

# After conversion (simplified)
def loop_test(x):
  return x > eps
def loop_body(x):
  return f(x)
x = ag.while_stmt(
  loop_test, loop_body, (x,))
\end{minted}
\end{listing}

The \code{for} statement is handled similarly. Similar to \code{if} statements, \code{while} and \code{for} loops may define symbols inside their body. If the loop body never executes, those symbols will remain undefined. This is also handled by using special ``undefined" values for the symbols that are not defined (as identified by liveness analysis) upon entry into the loop.

The overloaded control flow uses dynamic dispatch (see Appendix~\ref{appendix:features}).

\paragraph{Ternary Conditional Expressions}
The ternary operator \code{x if cond else y} is converted inline to the functional form \code{ag.if_stmt(cond, x, y)}.

\paragraph{Logical Expressions}
Binary and unary logical expressions can be handled using traditional operator overloading (e.g. \code{__lt__} for the \code{<} operator). However, \code{Tensor} does not support all operators for compatibility reasons (for example, \code{__eq__} is not supported). Therefore we replace certain binary and unary operators inline with overloadable functional forms. For example, \code{a and b} is replaced with \code{ag.and_(a, b)}.

\paragraph{Function Wrappers}
This conversion pass wraps the entire block of functions with additional boilerplate code. This accommodates for examples the necessary calls to create a TensorFlow \emph{name scope}, which improves the readability of the rendered graph. In addition, the function wrappers contain specialized error handlers that intercept certain errors to improve usability.

\section{Beyond TensorFlow: Alternate Back-Ends}
\label{lantern}
If TensorFlow is the only back-end of this code transformation, then the limitations of TensorFlow must also apply to AutoGraph. However, due to the nature of meta-programming, the SCT in AutoGraph can easily be used to target a variety of back-ends. As previously discussed, one shortcoming of TensorFlow is the inability to handle re-entrant in-graph functions, and by extension, recursive models. In order to showcase the utility of a general purpose SCT methodology as implemented by AutoGraph, we elect to target a new ML framework prototype called Lantern \cite{wang2018a, DBLP:journals/corr/abs-1803-10228}, which is capable of generating graphs describing recursive models.

\paragraph{The Lantern IR}
The Lantern back-end converts Lisp-like S-expressions describing numeric operations into efficient C++ code. Critically, Lantern supports programming features absent in the TensorFlow graph specification, like function recursion and in-line function definitions, which are essential in some state-of-the-art ML language models. We demonstrate the generality of AutoGraph by targeting the Lantern S-expression IR, which is supported by additional code conversion passes.

\paragraph{Staging Functions and Recursion}
In order to deal with functions in our model, we introduce two new functions: \code{__def_staging(function, *args)} and  \code{__call_staging(function, *args)}. These emit a function definition or call, respectively, in the generated S-Expression. Due to the deferred API presented by AutoGraph, we have the ability to specialize the generated functions in the S-Expression IR with respect to known parameters. Note that this specialization in function calls/definitions requires no additional modifications, as it is handled using the existing dispatching and overloading mechanisms present in AutoGraph. With the ability to define and call functions in the generated computation graph, this provides the interface necessary for defining and running recursive models.

To demonstrate this, we provide an end-to-end example of Python $\to$ S-Expr $\to$ C++. We first examine a recursive function in Python, as follows:

\begin{listing}[ht]
\begin{minted}[fontsize=\mintedsize]{python}
@ag.convert()
def tree_prod(base, tree):
    if not tree.is_empty:
        l = tree_prod(base, tree.left)
        r = tree_prod(base, tree.right)
        return l * r * tree.value
    else:
        return base
\end{minted}
\end{listing}

With the modifications in place which allow us to target Lantern, this will generate the following Python code (simplified for presentation):

\begin{listing}[ht]
\begin{minted}[fontsize=\mintedsize]{python}
def run(base, tree):
    def tree_prod(base, tree):
        def true_fn():
            return base

        def false_fn():
            l = __call_staged(tree_prod,
                    base, tree.left)
            r = __call_staged(tree_prod,
                    base, tree.right)
            return l * r * tree.value
        ag.if_stmt(tree.is_empty,
                true_fn, false_fn)
    __def_staged(tree_prod, base, tree)
    return __call_staged(tree_prod, base,
                         tree)
\end{minted}
\end{listing}

Note that in order to correctly generate the staged function, \code{__def_staged} must be passed the arguments which will eventually be passed to the function being defined. Running this generates S-Expression code, which is then fed as input to Lantern, which performs some internal computations and eventually generates and executes the following C++ code:



\begin{listing}[ht]
\begin{minted}[fontsize=\mintedsize]{C}
double Snippet(double base, Tree tree) {
  auto rec = [&](Tree tree,
  function<double(double)> cont,
  double base) {
    double grad = 0.0;
    if (!tree.is_empty) {
      auto cont_l = [&](double x1) {
        double sub_grad = 0.0;
        auto cont_r = [&](double x2) {
          double x3 = tree.value;
          double x4 = cont(x1 * x2 * x3);
          double x5 = x3 * x4;
          sub_grad += x2 * x5;
          return x1 * x5;
        };
        grad += rec(tree.R, cont_r, base);
        return sub_grad;
      };
      grad += rec(tree.L, cont_l, base);
    } else
      grad += cont(base);
    return grad;
  };
  return rec(tree,
    [&](auto x){return 1.0;}, base);
}
\end{minted}
\end{listing}

As shown, staging a recursive function requires that the generated C++ code also be recursive (as noted by the \code{rec} function). We note that the generated C++ code looks fairly complicated, due to the handling of back-propagation. Back-propagation is implemented via callbacks (seen as continuations, noted by \code{cont}, \code{cont_l}, and \code{cont_r} in the code), the details of which can be referenced in \citet{wang2018a, DBLP:journals/corr/abs-1803-10228}.

\section{Evaluation}
We tested the utility of AutoGraph on several axes. First, we asked whether AutoGraph could improve the readability of ML code that relied on data-dependent control flow without incurring a performance penalty. Second, we tested if AutoGraph could be used to move computation usually left outside of the TensorFlow graph, such as the entire training process of stochastic gradient descent (SGD), inside the graph IR. Third, we tested if AutoGraph could be used to produce performant code using features not supported in the TensorFlow graph by targeting alternative IRs. We also prepared additional samples of more complex algorithms, including Neural Model Translation with Attention, Sequence-to-sequence, MAML metalearning and L-BFGS optimizations. These can be found in Appendix~\ref{appendix:examples}.

\paragraph{RNN cells}
The code snippet below is an implementation of an RNN model that on simple inputs produces results identical to TensorFlow's built-in \code{tf.dynamic_rnn} function and runs at similar speed.

\begin{listing}[ht]
\begin{minted}[fontsize=\mintedsize]{python}
def dynamic_rnn(rnn_cell, input_data,
  initial_state, sequence_len=None):
  input_data = tf.transpose(input_data, 
    (1, 0, 2))
  outputs = []
  ag.set_element_type(outputs, tf.float32)
  state = initial_state
  if sequence_length is None:
    max_len = tf.shape(input_data)[0]
  else:
    max_len = tf.reduce_max(sequence_len)
  for i in tf.range(max_len):
    prev_state = state
    output, state = rnn_cell(input_data[i], 
      state)
    state = tf.where(
        i < sequence_len,
        state,
        prev_state)
    outputs.append(output)
  outputs = ag.stack(outputs)
  outputs = tf.transpose(outputs,
    (1, 0, 2))
  return outputs, state
\end{minted}
\end{listing}

Compare this terse and readable implementation to the equivalent graph version in Appendix~\ref{appendix:drnn}.


\begin{table}[htbp]
\centering
\caption{RNN Cell Performance (1K examples/sec)}
\resizebox{\columnwidth}{!}{%
\begin{tabular}{r|cccccc}
    \toprule
    Sequence Size &
    \multicolumn{3}{c}{Seq Size: 64} & 
    \multicolumn{3}{c}{Seq Size: 128} \\
    \cmidrule(lr){2-4}
    \cmidrule(lr){5-7}
    Batch Size & 32 & 64 & 128 & 32 & 64 & 128 \\
    \midrule
    Eager &
    $0.82 \pm 0.08$ & $1.57 \pm 0.13$ & $2.04 \pm 0.14$ & $0.43 \pm 0.03$ & $0.76 \pm 0.05$ & $1.04 \pm 0.06$ \\
    Official &
    $2.88 \pm 0.11$ & $3.63 \pm 0.13$ & $5.13 \pm 0.15$ & $1.44 \pm 0.04$ & $1.91 \pm 0.06$ & $2.61 \pm 0.05$ \\
    Handwritten &
    $2.95 \pm 0.13$ & $3.71 \pm 0.15$ & $5.24 \pm 0.11$ & $1.52 \pm 0.06$ & $1.96 \pm 0.07$ & $2.68 \pm 0.03$ \\
    AutoGraph &
    $2.72 \pm 0.09$ & $3.61 \pm 0.12$ & $5.05 \pm 0.10$ & $1.37 \pm 0.04$ & $1.86 \pm 0.06$ & $2.59 \pm 0.04$ \\
    \bottomrule
\end{tabular}
}
\end{table}

We compared TensorFlow's official implementation of \code{tf.dynamic_rnn} with both a hand-written, graph-based implementation, and the code snippet above converted into graphs via AutoGraph. Each run consisted of an execution of an RNN having hidden size 256, while varying batch sizes and the sequence length. Five warm-up runs were executed, and the mean and standard deviation of the 100 following runs are reported. For all examples, each run is executed as one \code{tf.Session.run()} call. All benchmarks were run on a dual-threaded 6-core Intel Xeon E5-1650 CPU. The use of AutoGraph improves the readability of the code and has a very minor effect on performance.

\paragraph{In-Graph Training}
Typically, a TensorFlow graph representing a single training step is executed repeatedly in a Python training loop outside of TensorFlow. This method is used because of the difficulty of using control flow operators within TensorFlow graphs, and incurs additional computational overhead. Here, we use AutoGraph to demonstrate a training loop that is implemented entirely as a computation graph. We trained a single linear layer on MNIST with stochastic gradient descent (SGD), and compared its performance with several other implementations. The first approach was TensorFlow Eager, an imperative execution mode for TensorFlow similar to NumPy and PyTorch. The second approach we tested was a traditional TensorFlow training process. The third approach was an in-graph training loop implemented using the TensorFlow \code{while_loop} API.

%


\begin{table}[htbp]
\centering
\caption{Model and Training Loop}
\begin{tabular}{r|c}
    \toprule
    & SGD Steps / sec \\
    \midrule
    Eager & $274.1 \pm 3.6$ \\
    Model In Graph, Loop In Python & $484.1 \pm 7.7$ \\
    Model And Loop In Graph & $646.5 \pm 14.1$ \\
    Model And Loop In AutoGraph & $623.5 \pm 13.5$ \\
\bottomrule
\end{tabular}
\end{table}

Each run consisted of 1000 training steps with a batch size of 200. One warm-up run was executed, and the mean and standard deviation of the 10 following runs are reported. For the in-graph training loop examples, the entire set of 1000 training steps is executed in one \code{tf.Session.run()} call. For the other examples, each training step is run as a separate \code{tf.Session.run()} call.
Executing a single-training-step graph repeatedly in a Python loop (the traditional approach) is faster than the eager-style code by 75\%. Moving the entire training process into a TensorFlow graph further yielded a roughly 30\% speedup.

\subsection{AutoGraph + Lantern: TreeLSTM}
\label{treelstm}
We evaluated a model of TreeLSTM for Sentiment Classification running on the dataset of the Stanford Sentiment \cite{D13-1170}, following the work of \cite{DBLP:journals/corr/TaiSM15}. The model embeds sentence parse-trees by recursively embedding the left/right sub-trees and combining the embedding vectors via BiLSTM core. The embedding of whole sentences is then passed to MLP for sentiment prediction. The model can be easily expressed in PyTorch using recursive functions, or in AutoGraph targeting recursive functions in Python. 
The final generated C++ code was compared against the PyTorch implementation in terms of training efficiency. To approximate a ``real-world'' running time, this experiment was run using a single thread on a laptop with a dual-core AMD A9-9410 Radeon CPU @ 1.70GHz and 8GB of SODIMM Synchronous 2400 MHz RAM, with Ubuntu 16.04.

Our AutoGraph implementation of TreeLSTM targeting Lantern yielded performance approximately 2.38 times faster than that of the PyTorch implementation. Our system achieved approximately 36.75 SGD steps per second, compared with the 15.41 steps per second using the PyTorch implementation. We note that we used a batch size of 1 for both systems due to difficulty in batching recursive models.


\begin{table}[htbp]
\centering
\caption{TreeLSTM Targeting Lantern}
\resizebox{\columnwidth}{!}{%
\begin{tabular}{r|c}
    \toprule Moved to separate files.
    & SGD Steps / sec \\
    \midrule
    Loop and Model in PyTorch & $15.41$ \\
    Loop and Model in AutoGraph/Lantern & $36.75$ \\
    \bottomrule
\end{tabular}
}
\end{table}

\section{Discussion}
\label{discussion}

Developing a source code transformation methodology is far from mechanical. There exist a number of design decisions which may ultimately yield different results in terms of expressiveness, performance and portability. In this section, we discuss some of these decisions and provide insight regarding how they shaped the current state of AutoGraph, including its current limitations. We provide a detailed discussion of error handling in AutoGraph in the Appendix~\ref{appendix:errors}.

\paragraph{Engineering Practices as a Feature}
The code conversion passes we implement in AutoGraph are non-local, and can interact with each other in complicated ways. For instance, converting deeply-nested \code{for} loops and \code{if} statements exposes dataflow interactions between each level of nesting. In order to build a reliable system, we made extensive use of engineering best-practices. For instance, all static analyses, code transforms, and utility functions are extensively unit tested ($>$50\% of the ~22k LOC in AutoGraph is tests). Further, interactions between features are tested in end-to-end reference tests. Any changes to the AutoGraph system require that all unit and reference tests pass, and all code is manually reviewed by at least one engineer for correctness, readability and adherence to style guidelines. Anecdotally, this test- and review-oriented development practice has caught many surprising and subtle bugs, and allowed a library as complex as AutoGraph to remain relatively easy to maintain and extend. Further, we built many useful utilities for manipulating Python source code that simplified development (described in Appendix~\ref{appendix:utils}).

\paragraph{Alternative Approaches for Implementing Staged Programming}
An alternative approach to SCT would have been to build a new Python interpreter with non-standard execution semantics for Python programs that could map to TensorFlow graphs, and indeed, an early proposal for AutoGraph was to do exactly this. However, a non-standard Python interpreter would require reimplementing all aspects of the Python language, including those parts that require no modifications in machine learning code. 

We could also parse Python to our own intermediate representation, a strategy taken recently by the Myia system \cite{merrienboer2018myia}. This intermediate representation could then be either back-converted to Python or executed in a dedicated VM. Indeed, this strategy is similar to our ability to work with Lantern; AutoGraph modifies the original Python source code such that it generates S-Expressions as an IR, which are then consumed by Lantern.

Our choice to emit Python code after conversion has several advantages. Unsupported code idioms are allowed to pass through conversion if they do not affect the program semantics. This simplifies the support for legacy TensorFlow code. Further, the generated code can be inspected, and even modified by the user.

\paragraph{Comparing Torch Script and AutoGraph}
\label{torchscript_ag}
Similar to ONNX, PyTorch's Torch Script framework \cite{torchscript2018} allows users to save models for later evaluation, while providing an even higher-level interface for programming: nearly native Python with two new decorators. These decorators, \code{torch.jit.trace} and \code{torch.jit.script}, produce Torch Script code (a subset of Python used as an IR for the eventual computation graph) from idiomatic Python, though they accomplish this via different methods.

The \code{torch.jit.trace} decorator works as the name suggests: it extracts computation graphs through tracing. This produces fully shape-specialized Torch Script code, which allows for highly optimized models (and an easy target for potential compilers). However, tracing in Torch Script has the same drawback as found in ONNX: as stated clearly by the Torch Script developers, ``Tracing only correctly records functions and modules which are not data dependent (e.g., have conditionals on data in tensors)...''

Torch Script's \texttt{torch.jit.script} decorator, on the other hand, will directly translate the decorated Python function to Torch Script code, which does allow for data-dependent control flow. While this seems similar to AutoGraph's source code transformation model (detailed in Section~\ref{dispatch}), there are a number of important differences between the two methodologies. Torch Script is inherently bound to the PyTorch runtime, which prevents the use of Torch Script with any other specialized or accelerated ML back-ends. Furthermore, \code{torch.jit.script} does all of its work at compile time, and thus the only view of staging available currently is the ability to do shape propagation on a dynamically-shaped graph (resulting from \code{torch.jit.script}). This drawback comes as a result of the decision to target a relatively basic IR (Torch Script), rather than Python code. One powerful consequence of this decision, however, is the ability to cleanly implement autobatching on Torch Script, which is otherwise difficult in systems targeting a broader IR.

\paragraph{Limitations}
The Python language is large, and AutoGraph does not stage all of it. We focus on the subset that enables machine learning programming, but we are still missing many useful constructs, such as associative data structures and \code{try/except} blocks. In some cases, there is no corresponding construct in the TensorFlow or Lantern IR, but as we build support for more IRs, we anticipate being able to successfully convert more of the Python language. Although only a subset of the Python language is converted to TensorFlow constructs, AutoGraph does allow nearly every Python construct, and will simply call it unconverted. This allows AutoGraph to be compatible with the vast majority of existing graph code. Appendix~\ref{appendix:features} exhaustively documents Python language support in AutoGraph.

In addition, the data-dependent staging decisions made by AutoGraph are obscured from the user, much like Python operator overloading obscures computation made in the overloaded operators. For instance, if the user accidentally passes a Python boolean instead of a TensorFlow boolean to a conditional, it will not be staged into a graph, with potential performance implications. Currently, the user has few tools to catch and debug this behavior. We already provide better error messages than a system like this naively would (see Appendix~\ref{appendix:errors}), but further work is required.

Additional challenges arise from the mismatch between Python and the IRs typing system. For example, TensorFlow does not support nullable types, so we impose additional constraints on the Python semantics by requiring that all code paths initialize a variable when control flow is staged in TensorFlow. Similarly, because Python types like lists are generic, element access lacks type information and we may require additional user annotations when the IR is strongly typed, which is usually the case. More advanced type inference mechanics that could obviate these annotation is a subject for future work.

We make a best effort to guarantee that the conversion to IR is either semantics-preserving, or it explicitly fails. However, a more rigorous treatment of the correctness of our system is needed. We plan to treat this both formally and empirically, using a random code generation fuzzing system. In the meantime, we provide as evidence of correctness an expansive test suite for AutoGraph, containing hundreds of tests. Furthermore, due to AutoGraph being included in \code{tf.function}, the default way to accelerate code in TensorFlow 2.0, AutoGraph is also subject to all tests covering the TensorFlow codebase. While this notion of test-based correctness does not provide a formal guarantee of correctness, we note that this is consistent with other formal analyses of Python semantics~\cite{DBLP:conf/oopsla/PolitzMMWPLCK13}.

Lastly, AutoGraph relies on Python introspection and reflection APIs, such as \code{inspect} and \code{imp}. While these are available in the vast majority if use cases, there are instances when AutoGraph cannot be used, for example when source code information is not available.

\section{Conclusions and Future Work}

We have described AutoGraph, a staged programming system for automatically rewriting idiomatic Python code into an equivalent lower-level IR, including TensorFlow graphs and other, more experimental, back-ends. AutoGraph achieves a balance in the design space between {\em imperative} and {\em graph-based} code. These two programming models -- fully-imperative with high runtime overhead, and fully-staged with high developer mental overhead -- are not binary choices. Using SCT, we can eliminate the distinction between the two. We believe that this approach is applicable broadly, and are working to target a wider suite of IRs in new applications.

The entirety of AutoGraph is open sourced via the TensorFlow project on GitHub  at \url{https://github.com/tensorflow/tensorflow/tree/master/tensorflow/python/autograph}.


\section*{Acknowledgements}
We would like to thank Alex Passos and the rest of the TensorFlow team for their help and support integrating AutoGraph into TensorFlow 2.0.

The technique based on dynamic dispatch was studied in prior work by Josh Levenberg.

\bibliography{main}
\bibliographystyle{sysml2019}

\clearpage
\appendix

\section{Dynamic RNN Implementation}
\label{appendix:drnn}
Below is the hand-written graph implementation of the \code{tf.dynamic_rnn} cell. 

\vspace{2ex}
\begin{minted}[fontsize=\footnotesize]{python}
def dynamic_rnn(rnn_cell, input_data,
  initial_state, sequence_len=None):
  input_data = tf.transpose(input_data,
    (1, 0, 2))
  outputs = tf.TensorArray(
      tf.float32, size=0, dynamic_size=True)
  if sequence_length is None:
    max_len = input_data.shape[0]
  else:
    max_len = tf.reduce_max(sequence_len)
  def while_body(i, state, outputs):
    prev_state = state
    output, state = rnn_cell(
        input_data[i], state)
    state = tf.where(
        i < sequence_len,
        state,
        prev_state)
    outputs = outputs.write(i, output)
    return i + 1, state, outputs
  def while_cond(i, state, outputs):
    return i < max_len
  _, state, outputs = tf.while_loop(
      while_cond,
      while_body,
      loop_vars=(tf.constant(0),
                 initial_state,
                 outputs))
  outputs = outputs.stack()
  outputs = tf.transpose(outputs, (1, 0, 2))
  return outputs, state
\end{minted}
\vspace{2ex}

\section{Error Handling}
\label{appendix:errors}

In AutoGraph, there are three distinct steps of execution in addition to the usual syntax verification performed by the Python runtime:

\begin{itemize}
    \item Conversion
    \item Staging (e.g., TensorFlow graph construction)
    \item Runtime (e.g., TensorFlow graph execution)
\end{itemize}

The latter two steps can be associated with the two stages in the multi-stage programming model that platforms like TensorFlow and PyTorch's JIT model implement.
Each of these steps has distinct requirements for error handling, but principally make use of these two technologies:

\begin{itemize}
  \item {\em Source map construction}. Each node in the AST, even after several passes of SCT, is associated to an original line of the user's Python code.
  \item {\em Error rewriting}. Several frames in the stack trace of TensorFlow code, especailly AutoGraph-generated TensorFlow code, point to lines of code written by the AutoGraph compiler system rather than the user. We are able to reassociate temporary files (used when generating code in AutoGraph) to the user's original source files.
\end{itemize}

\paragraph{Conversion Errors}

Conversion errors may occur due to code that is otherwise legal Python, but is unsupported by AutoGraph. These errors usually originate inside AutoGraph internal code.

For usability, such errors must indicate the location in the converted code of the idiom that caused the error. In addition, the error message must provide sufficient information to allow the developer to remedy the error. Lastly, the error stack trace should avoid references to internal code, as they are typically uninformative to the user.

Currently, we facilitate this requirement by generating a stack-trace-like message that indicates the location of the error. In the future, we plan to further improve the conciseness of error messages of this type.

\paragraph{Staging Errors}

Staging errors can occur in successfully converted code and are typically raised because of disallowed or invalid argument types, shapes, hyperparameter values or other conditions that are only detectable at runtime. To address this, we plan to generate a stack-trace-like message with frames from the original code from which the intermediate code was generated. This is facilitated by the AST source map that we maintain between each node in the generated AST and the user's original source code.

Another challenge is that error messages may refer to generated symbols or to contexts specific to generated code. Addressing this shortcoming is a subject of future work.

\paragraph{Runtime Errors}

The name of this class of errors refers to the staged IR runtime.

For example, integer division by zero errors in TensorFlow:

\vspace{-2ex}
\begin{listing}[ht]
\begin{minted}[fontsize=\footnotesize]{python}
def f(n):
  return tf.constant(10, dtype=tf.int32) / n
\end{minted}
\end{listing}
\vspace{-5ex}

The IR execution environment typically includes facilities to trace the source of the error to user code. However, in the case of AutoGraph, that will be generated code. To remedy this, we plan to intercept these errors and attach information that helps the user further trace the source of the error to original, pre-conversion code. We plan to enhance the user experience with the addition of \code{tf.function} in the TensorFlow 2.0 API.

\section{Useful Utilities}
\label{appendix:utils}
 

In order to build the system as described, we created a large library of source code transformation tools that we anticipate will be useful to the broader Python community.

\paragraph{Easy Code Quoting and Unquoting}


A few of the utility functions are listed below:

\begin{itemize}
  \item \code{parser.parse_entity(fn_or_class)} takes a Python class or function and returns the corresponding AST node, wrapped in a containing \code{Module} node.
  \item \code{parser.parse_str(code_string)} is dentical to \code{parse_entity}, except takes a string of Python code as input. The string may contain any valid Python code.
  \item \code{pretty_printer.fmt(ast_node)} returns a pretty-printable string representing the AST.
  \item \code{compiler.ast_to_source(ast_node)} unparses an AST into the equivalent Python code, returned as a string.
  \item \code{compiler.ast_to_object(ast_node)} compiles an AST into an equivalent Python entity, returned as a module.
\end{itemize}

For example:

\vspace{-2ex}
\begin{listing}[ht]
\begin{minted}[fontsize=\footnotesize]{python}
node = parse_str('a = b')
print(fmt(node))

# Output:
Module:
| body=[
| | Assign:
| | | targets=[
| | | | Name:
| | | | | id="a"
| | | | | ctx=Store()
| | | | | annotation=None
| | | ]
| | | value=Name:
| | | | id="b"
| | | | ctx=Load()
| | | | annotation=None
| ]
\end{minted}
\end{listing}
\vspace{-5ex}

These utilities make it easy to make small modifications to the AST.

\vspace{-2ex}
\begin{listing}[ht]
\begin{minted}[fontsize=\footnotesize]{python}
node = parse_str('a = b')
node.body[0].value.id = 'c'
print(ast_to_source(node))

# Output:
a = c
\end{minted}
\end{listing}
\vspace{-5ex}

\paragraph{Templated Code Rewriting}
Example:

\vspace{2ex}
\begin{minted}[fontsize=\footnotesize]{python}
code_quote = '''
def fn(args):
  body
'''
new_body = textwrap.dedent('''
  a = x
  b = y
  return a + b
''')
node = templates.replace(
  code_quote,
  fn='my_function',
  args=('x', 'y'),
  body=parser.parse_str(new_body).body
)
print(compiler.ast_to_source(node))

# Output:
def my_function(x, y):
  a = x
  b = y
  return a + b
\end{minted}
\vspace{2ex}

The function inserts string symbols or AST nodes into the quoted code template, and performs additional integrity checks. This allows for the easy construction of complicated code blocks, especially with respect to building the AST manually.

\section{Expanded Examples}
\label{appendix:examples}
We expand on the toy examples in the main paper to illustrate AutoGraph's utility when implementing more realistic algorithms and models. These were implemented using TensorFlow's benchmark utilities\footnote{\url{https://www.tensorflow.org/community/benchmarks}} so that they can more easily be run. This also allows us to compare the performance of AutoGraph generated code to other reference implementations both from AutoGraph's authors and distributed as part of TensorFlow. We report some preliminary findings for each example.

All example code mentioned in this section, as well as the full runnable code for examples found through the paper, can be found at \url{https://github.com/tensorflow/autograph/examples/sysml2019}.

\subsection{Beam Search}
Beam search is an algorithm often used in machine translation. The algorithm builds candidate sequences by taking the most-likely steps at each transition, possibly discarding less-likely sequences. This is an interesting use-case for AutoGraph because beam search consists of complex computation and decisions at each step, with the number of steps capped at a the maximum sequence size. The simplest implementation of beam search is a loop that breaks if all candidate sequences have terminated. More robust implementations will separately keep track of living and terminal candidate sequences, and break if no living candidate has the potential to outscore a terminal candidate. Breaking out of the loop is essential to the performance of beam search since it often can generate sequences that are far shorter than the maximum allowable size.

We implemented beam search using TensorFlow Eager. Using AutoGraph, the benchmark runs between 2 and 3.2 times faster than the same code run using TensorFlow Eager. The improvement varies as we change the maximum sequence length and vocabulary size. Longer sequences and smaller vocabularies typically show more improvement when using AutoGraph. Longer sequences result in more iterations of the loop, so embedding these loops in the TensorFlow graph with AutoGraph shows more relative improvement. A larger vocabulary results in more expensive vector and matrix operations, taking longer overall.

\subsection{L-BFGS}
The L-BFGS (Limited-Memory Broyden–Fletcher–Goldfarb–Shannon) algorithm is often used for parameter estimation in Machine Learning. Our implementation is based on the TensorFlow Eager implementation written by Yaroslav Bulatov\footnote{\url{https://github.com/yaroslavvb/stuff/tree/master/eager_lbfgs}}. In our benchmark, AutoGraph is almost 2 times faster than Eager with a batch size of 10 in approximately the same amount of code.

\subsection{Model-Agnostic Meta-Learning (MAML)}
Model-Agnostic Meta-Learning (MAML, \citet{DBLP:conf/icml/FinnAL17}) is an algorithm for meta-learning, especially effective for few-shot learning. Our benchmark is based on the sinusoidal example from \citet{DBLP:conf/icml/FinnAL17}.\footnote{\url{https://github.com/cbfinn/maml}}

We implemented our MAML benchmark using code that is compatible with both TensorFlow Eager AutoGraph. When training a single meta-parameter, the AutoGraph converted code ran 1.9 times faster than the identical code run in Eager mode. AutoGraph converted code was 2.7 times faster when training 10 meta-parameters.

\subsection{seq2seq}

The seq2seq (Sequence-to-Sequence) model\footnote{\url{https://google.github.io/seq2seq/}} is a general purpose encoder and decoder that can be used for tasks like machine translation. We implemented this model and a benchmark that measures the performance of the model on random input sequences. 

We implemented this benchmark in TensorFlow Eager and converted that Eager code using AutoGraph. AutoGraph converted code was 1.18 to 3.05 times faster than the Eager equivalent. The performance improvement varies with vocabulary size: AutoGraph performs better on larger vocabularies. Varying sequence length from 64 to 128 had minimal effect on the performance improvement. We also implemented optional ``teacher forcing", which almost doubles the improvement gained from AutoGraph. This is because teacher-forcing reduces the amount of time spent performing computations, so the overhead of Eager mode is a larger percentage of the overall time. AutoGraph is designed to reduce such overhead, in this case by embedding data-dependent control flow in the graph executed by TensorFlow.

\section{Supported Features}
\label{appendix:features}
Tables \ref{tbl:features1}, \ref{tbl:features2} and \ref{tbl:features3} show the features of Python and TensorFlow that are presently supported by AutoGraph.

\begin{sidewaystable*}
\centering\small
\begin{tabular}{@{}l|l|p{5cm}|p{5cm}|p{5cm}}
  \multicolumn{2}{ c| }{} & Conversion Triggers & Python Semantics & TensorFlow Semantics

\\ \hline
Control Flow
& if 
& cond is \texttt{Tensor}-like \emph{or} a nest collection\footnote{A ``nest collection'' is a collection that is recognized by \texttt{tf.nest.}} of \texttt{Tensor}-like 
& extraneous side effects possible when object mutation is used\footnote{Example: conditionally setting an attribute / item may be changed to always set that attribute / item. This is something we plan to remedy in TF 2 release.} 
& \texttt{tf.cond}.
Object mutation inside control flow limited to visible operations\footnote{For example: attribute and item mutations done inside the control flow body preserve semantics. Mutations done in functions calls do not necessarily preserve semantics.}
all code paths must produce consistent value

\\\cline{2-5}
& for
& iterated is \texttt{Tensor}-like \emph{or}
iterated is nest collection of \texttt{Tensor}-like \emph{or}
iterated is \texttt{tf.Dataset} \emph{or}
iterated is \texttt{tf.QueueBase}
& extraneous side effects possible when object mutation is used
& \texttt{tf.while\_loop} for \texttt{Tensor} and \texttt{tf.QueueBase}
\texttt{Dataset.reduce} for \texttt{tf.Dataset}. All code paths must produce consistent value

\\\cline{2-5}
& while
& condition \emph{closure} is collection of any \texttt{Tensor}-like\footnote{When staging while loops using \texttt{tf.while\_loop}, the condition of the loop is only evaluated by \texttt{tf.while\_loop} itself. However, we need to determine whether the loop is will be staged or not \emph{before} \texttt{tf.while\_loop} is called. For this reason, we do not evaluate the loop condition beforehand to avoid causing any Python side effects triggered by the evaluation of the loop condition twice. In the future, we plan to evaluate the condition function twice and clearly document this semantic.}
& preserved
& \texttt{tf.while\_loop}, all code paths must produce consistent value

\\\cline{2-5}& break / continue / return
& lowered to conditional checks  
& preserved
& \texttt{tf.cond}, all return values must have a consistent value.

\\\cline{2-5}
& try / except / finally / raise
& not converted\footnote{There is no currently no support for catching exceptions in TensorFlow.} 
& preserved 
& n/a

\\\cline{2-5}
& with
& not converted
& preserved
& n/a

\\\cline{2-5}
& yield
& not allowed\footnote{Plans to support \texttt{yield} (without conversion) soon.}
& n/a
& n/a

\\ \hline
Operators
& unary
& argument is \texttt{Tensor}-like
& preserved
& corresponding TF op

\\\cline{2-5}
& binary, arithmetic
& not converted \footnote{Note however that \texttt{Tensor} objects typically overload all arithmetic operators and expressions will be staged into TF ops.}
& preserved
& corresponding TF op

\\\cline{2-5}
& binary, equality
& either argument is \texttt{Tensor}-like
& preserved
& corresponding TF op

\\\cline{2-5}
& binary, boolean
& either argument is \texttt{Tensor}-like
& preserved
& lazy boolean\footnote{For example, \texttt{x and y} is converted to \texttt{tf.cond(x, lambda: y, lambda: x)} to be consistent with Python’s lazy boolean evaluation semantics.} using \texttt{tf.cond}

\\\cline{2-5}
& ternary conditional
& either argument is \texttt{Tensor}-like
& preserved
& \texttt{tf.cond}

\end{tabular}
\caption{AutoGraph Supported Features}
\label{tbl:features1}
\end{sidewaystable*}

\begin{sidewaystable*}
\centering\small
\begin{tabular}{@{}l|l|p{5cm}|p{5cm}|p{5cm}}
  \multicolumn{2}{ c| }{} & Conversion Triggers & Python Semantics & TensorFlow Semantics

\\ \hline
Functions
& user-defined
& (converted in recursive mode \emph{and} not part of a whitelisted module\footnote{Currently, the whitelist includes the TF module.}) \emph{or} function is passed directly to AutoGraph\footnote{That is, user functions directly passed to \texttt{to\_graph} or \texttt{tf.function} are always converted.}
& preserved
& inlined

\\\cline{2-5}
& lambda
& converted in recursive mode
& preserved
& inlined

\\\cline{2-5}
& constructors
& not converted
& preserved
& n/a

\\\cline{2-5}
& instance methods
& converted
& output is unbound function\footnote{That is, it is a function that takes \texttt{self} as first argument.}
& inlined

\\\cline{2-5}
& class methods
& converted
& output is unbound function
& inlined

\\\cline{2-5}
& built-in
& converted: \texttt{print}, \texttt{len}, \texttt{range}, \texttt{int}, \texttt{float}
& preserved
& corresponding TF op\footnote{Not all Python built-ins have a corresponding TF op.}

\\\cline{2-5}
& native
& not converted
& preserved
& n/a

\\ \hline
Collections
& list literals
& \textbf{experimental}; list is empty
& preserved
& low level \texttt{Tensor} list

\\\cline{2-5}
& list append
& target is low-level \texttt{Tensor} list \em{or} target is \texttt{tf.TensorArray}
& preserved
& low-level \texttt{Tensor} list push back, \texttt{tf.TensorArray.write} respectively

\\\cline{2-5}
& list pop
& target is low-level \texttt{Tensor} list
& preserved
& low-level \texttt{Tensor} list pop back

\\\cline{2-5}
& other (\texttt{dict}, \texttt{set}, etc.)
& not converted\footnote{Plans to add support as corresponding TF ops are added.}
& preserved
& n/a

\\\cline{2-5}
& get item / set item
& target is \texttt{Tensor} \em{or} target is \texttt{tf.TensorArray} \em{or} target is low-level \texttt{Tensor} list
& preserved
& \texttt{Tensor.\_\_getitem\_\_} / \texttt{Tensor.\_\_setitem\_\_}, \texttt{tf.TensorArray.read} / \texttt{tf.TensorArray.write}, low level \texttt{Tensor} list get item / set item respectively

\\ \hline
Comprehensions
& 
& not converted
& preserved
& n/a

\end{tabular}
\caption{AutoGraph Supported Features (continued)}
\label{tbl:features2}
\end{sidewaystable*}

\begin{sidewaystable*}
\centering\small
\begin{tabular}{@{}l|l|p{5cm}|p{5cm}|p{5cm}}
  \multicolumn{2}{ c| }{} & Conversion Triggers & Python Semantics & TensorFlow Semantics

\\ \hline
Variables
& undefined
& 
& reified with special value\footnote{Long term plans to fully obey Python semantics and raise a runtime exception when undefined variables are accessed.}
& not allowed

\\\cline{2-5}
& global
& not allowed\footnote{Plans to support it soon.}
& n/a
& n/a

\\\cline{2-5}
& nonlocal
& not allowed\footnote{Plans to support it soon.}
& n/a
& n/a

\\ \hline
Literals
& 
& not converted
& preserved
& n/a\footnote{Note that many TF ops autobox certain values to \texttt{Tensor}.}

\\ \hline
Classes
& class types
& \textbf{experimental}; class is passed directly to AutoGraph\footnote{That is, user classes directly passed to \texttt{to\_graph} or \texttt{tf.function} are always converted.}
& preserved
& new class with all qualifying\footnote{See User functions. For example, a subclass of a Keras Model class will only convert the methods defined in the subclass, not the methods inherited from the Model class.} methods converted

\\\cline{2-5}
& objects
& object is callable
& preserved
& object's \texttt{\_\_call\_\_} is converted

\\\cline{2-5}
& get / set attribute
& not converted
& preserved
& n/a

\\ \hline
Decorators
& user
& converted in recursive mode
& preserved
& n/a

\\\cline{2-5}
& built-in
& not converted; some not allowed\footnote{Examples: \texttt{functools.lru\_cache} is not supported. \texttt{functools.wraps} is supported, but not converted.}
& preserved
& n/a

\\ \hline
Generators
& 
& not allowed\footnote{Plans to allow the use of generators, without conversion.}
& preserved
& n/a

\\ \hline
Power Features
& exec
& not supported
& n/a
& n/a

\\\cline{2-5}
& pdb
& not converted
& partially\footnote{The \texttt{pdb} calls will be inserted in the generated code, and will take effect at staging, when the graph is constructed.}
& n/a

\\\cline{2-5}
& inspect
& not converted
& partially\footnote{Some \texttt{inspect} APIs, like \texttt{getsource} work correctly, but we have not extensively tested them.}
& n/a

\end{tabular}
\caption{AutoGraph Supported Features (continued)}
\label{tbl:features3}
\end{sidewaystable*}


\end{document}